# An algebraic Approach to $Z$-Factors

Dirk Kreimer[†]

*Department of Physics, University of Tasmania, GPO Box 252C*
*Hobart, Tasmania 7001, Australia*

ABSTRACT

We sketch a systematic approach to multiloop calculations. We focus on the $Z$-factors of a renormalizable theory and show that they can be obtained by purely algebraic methods based on powercounting and the forest structure of the divergent graphs.

## 1. Introduction

It is the purpose of this paper to sketch a systematic approach to multi-loop calculations, especially in the Standard Model, where one is confronted with various different masses in the processes. We will be concerned with the $Z$-factors of the theory, that is our aim is a quick and systematic approach to the renormalization of the Standard Model.

The main result is the following: The two-loop renormalization of the Standard Model can be obtained by algebraic methods based on powercounting and kwowledge of the topology of the considered graphs. We will only sketch the generalization of this result to the multi-loop case here.

We will demonstrate the idea in the following by discussing an easy two-loop two-point function. Our presentation is close to our previous work.[1] There it was shown how one can separate the UV divergent parts of a given Feynman graph in a systematic manner to obtain characteristic integral representations for the graph as a whole, plus some analytic results containing its ultraviolet divergences. Here we go one step further and, inspired by these previous results, show how to extract the renormalization constants algebraically. The crucial ingredient is the observation that the calculation of the one-loop counterterm subgraphs can be embodied in a way which simplifies the whole calculation instead of being an additional amount of work.

Indeed, a short glance at the results of renormalization theory tells us that this is not a very unlikely result.

---

## 2. An Example

As an example we consider Fig.(1) where all particles and vertices are the ones of a renormalizable theory.

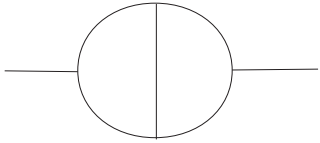

Fig. 1. The example.

Consider the following table.

Table 1. The structure of the UV-divergences.

| Graph | $\sim \frac{1}{(D-4)^2}$ | $\sim \frac{1}{(D-4)}$ | finite |
|---|---|---|---|
| | local | nonlocal | nonlocal |
| | local | nonlocal | nonlocal |
| | local | nonlocal | nonlocal |
| $\sum$ | local | local | nonlocal |

In the left column we have listed the graph of Fig.(1) and its counterterm graphs. The other columns specify the nature of the various coefficients of the Laurent expansion in $(D-4)$. Typically, the leading divergence has only polynomial behaviour in exterior momenta, i.e. local coefficients, while the subleading divergence and the finite part provide non-local functions. As we are considering a renormalizable theory, we see that in the sum of the graph plus its counterterm graphs, we have a local coefficient also for the subleading divergence.

It is the very merit of a renormalizable theory to allow for such local Z-factors. This indicates that it might be easier to obtain the full Z-factor (at least its divergent part, that is its value in the MS scheme) than the results for the graph and the counterterm graphs separately.

To this end let us have a closer look at Fig.(1). It has, as depicted in the second and third row of Tab.(1) two divergent subgraphs. These give rise to vertex corrections. Assuming these vertices $\Gamma$ to be derived from a renormalizable lagrangian we would have Z-factors multiplying the vertices: $\Gamma \rightarrow Z\Gamma$. In dimensional regulariza-tion the $Z$-factor (in one-loop order) has the form

$$Z^{(1)} = \frac{1}{D-4}z_{-1} + z_0,\tag{1}$$

with some constants $z_{-1}, z_0$. In particular, these constants are independent of the momentum transfer at the vertex. In the MS-scheme we have $z_0 = 0$. Usually one obtains these $Z$-factors by calculating the corresponding one-loop graph and establishes some renormalization condition on it. Here we invert this program and reexpress the $Z$-factor through its corresponding graph:

$$Z^{(1)} = \left( {}_q \!\!-\!\!\!\prec\!\!\!\triangleleft \right) \Big|_{\substack{q=0 \\ m_i=0}} (k^2)^{(4-D)/2} - c_0 + z_0.\tag{2}$$

So we evaluate the graph at momentum transfer zero, -$q = 0$-, and have set all internal masses to zero. The result is a function of a momentum $k$ say which scales with a power given by $(k^2)^{(D-4)/2}$, due to the fact that the vertex has a dimensionless coupling. Cancelling the scaling by multiplying with the inverse power, we obtain a Laurent series in $(D-4)$. We subtract the finite coefficient of this Laurent series, $c_0$, and add the finite value given by the chosen renormalization prescription. We stress once more that the $Z$-factors are independent of the momentum transfer so that we could also establish the above identity at non-vanishing momentum transfer, in which case we would have $c_0 \equiv c_0(q,k)$.

Now let us make use of this identity in the second and third column of Tab.(1). We define $d_0 := -c_0 + z_0$ and obtain

$$\begin{aligned}
-\!\!\bigcirc\!\!- &= \left( -\!\!\ominus\!\!- \right)\Big|_{\substack{q=0 \\ m_i=0}} (k^2)^{(4-D)/2} + d_0 -\!\!\bigcirc\!\!-\,, \\
-\!\!\bigcirc\!\!- &= \left( -\!\!\ominus\!\!- \right)\Big|_{\substack{q=0 \\ m_j=0}} (l^2)^{(4-D)/2} + d_0 -\!\!\bigcirc\!\!-\,,
\end{aligned}\tag{3}$$

where we assumed that $m_i, l$ are the masses and loop momentum corresponding to the one subdivergence and $m_j, k$ correspond to the other subdivergence. Here, the first line has to be integrated in $l$ first, the second one in $k$ first.

According to what we have said in the introduction we expect some simplifications when we add these two terms to the original graph. The above identities show that the expressions obtained after performing one loop integration in the original graph are almost the same as the ones corresponding to the counterterm graphs. Here we use the fact that the original graph can be written as:[1]

$$\begin{aligned}
-\!\!\ominus\!\!- = &\left( -\!\!\ominus\!\!- \; - \; -\!\!\ominus\!\!-\Big|_{\substack{q=0 \\ m_i=0}} \; - \; -\!\!\ominus\!\!-\Big|_{\substack{q=0 \\ m_j=0}} \; + \; -\!\!\ominus\!\!-\Big|_{\substack{q=0 \\ m_i=m_j=0}} \right) \\
&+ \; -\!\!\ominus\!\!-\Big|_{\substack{q=0 \\ m_i=0}} \; + \; -\!\!\ominus\!\!-\Big|_{\substack{q=0 \\ m_j=0}}\,.
\end{aligned}\tag{4}$$

The term in the bracket on the right-hand side does not involve any UV-divergences which result from the subdivergences, and is thus free of nonlocal terms, while the other two terms almost equal the expressions generated by the subdivergences, as can be seen by comparing them with Eq.(3).

In particular, also in the spirit of the approach cited above, we see that the main difference between these expressions are factors of the type $(l^2)^{(4-D)/2}$ or $(k^2)^{(4-D)/2}$. Were we allowed to set $D = 4$ in the above expressions, the powers of $l^2$ resp. $k^2$ would vanish. Direct cancellations between the original graph and its counterterm graphs would then occur also in the subleading ($\sim \frac{1}{D-4}$) UV-divergent sector. Unfortunately there are further UV-divergences involved from the remaining loop integration. But we can isolate these UV-divergences in very simple integrals. The remaining terms contain all the dependence on masses and exterior momenta but are UV finite. So they will cancel and thus not contribute to our MS-scheme $Z$-factors.

As it was shown elsewhere not all the terms involved in a two-loop calculation give rise to 'real' two-loop integrals.[1] So before we start our calculation we would like to separate these terms, which are mainly 'squared one-loop contributions'. These contributions appear whenever numerator terms cancel the propagators carrying both loop momenta ($P_3$ say in our example):[1]

$$\frac{v + wP_3}{N_l P_3 N_k} \quad \rightarrow \quad \frac{v}{N_l P_3 N_k} + \frac{w}{N_l N_k},$$

$$\text{---}\bigoplus\text{---} \quad \rightarrow \quad \text{---}\bigoplus\text{---}\Big|_2 + \text{---}\bigcirc\bigcirc\text{---}, \tag{5}$$

where $|_2$ denotes the true two-loop contributions, and $v, w$ the corresponding expressions for the numerator, that is polynomials in the exterior momentum, loop momenta and masses, as determined by the chosen renormalizable field theory.

These squared one-loop terms are independent of the other loop momentum and have corresponding terms in the counterterm graphs. So, when adding the original graph to the counterterm graph they cancel out in an obvious identity:

$$\text{---}\bigoplus\text{---}\Big|_{q=0} + \text{---}\bigcirc\text{---}\Big|_{q=0} + \text{---}\bigcirc\text{---}\Big|_{q=0} =$$

$$\text{---}\bigoplus\text{---}\Big|_2\Big|_{q=0} + \text{---}\bigcirc\text{---}\Big|_2\Big|_{q=0} + \text{---}\bigcirc\text{---}\Big|_2\Big|_{q=0}, \tag{6}$$

which can be derived by using Eq.(2,3,5). The fact that the two loop momenta do not interfere can be concluded from the one-particle reducible structure of this term in the above graphical representation Eq.(5). So in the following it is always understood that we restrict ourselve to the remaining true two-loop terms. The terms generated by the counterterm graphs which are proportional to $d_0$ in Eq.(3) are one-loop graphs. So their UV-divergences can be calculated by making use of the fundamental identities

$$\int \frac{d^D k}{(k^2)^{\lambda_1}((q-k)^2)^{\lambda_2}} = \pi^D \frac{1}{(q^2)^{\lambda_1 + \lambda_2 - D/2}}$$

$$\frac{\Gamma(\lambda_1 + \lambda_2 - D/2)\Gamma(-\lambda_1 + D/2)\Gamma(-\lambda_2 + D/2)}{\Gamma(\lambda_1)\Gamma(\lambda_2)\Gamma(D - \lambda_1 - \lambda_2)}, \tag{7}$$

$$\int \frac{d^D k}{(k^2)^{\lambda_1}(k^2 + m^2)^{\lambda_2}} = \pi^D \frac{1}{(m^2)^{\lambda_1 + \lambda_2 - D/2}}$$
$$\frac{\Gamma(\lambda_1 + \lambda_2 - D/2)\Gamma(-\lambda_1 + D/2)}{\Gamma(\lambda_2)\Gamma(D/2)}. \tag{8}$$

Our conventions follow the book by Smirnov.[2]

We will now show that also the true two-loop contributions can be obtained by knowledge of this identity, as far as their UV-divergences are concerned.

To this end we treat all three graphs in a similar manner. The idea is to shift the UV-divergence of the first loop momentum to the second one plus some terms which again do not mix in both loop momenta. These terms will cancel in the sum of the three graphs, while the shifted terms are UV-convergent in the first loop momentum. We then isolate the UV-divergences of the second loop momentum in integrals evaluated for vanishing exterior momenta, which are of the form Eq.(7,8). The shifting of the UV-divergences of the first integration can be done for example by partial integrations (a well-known technique, introduced by K. Chetyrkin et.al.[3]).[1] Typically one generates contributions of the form

$$0 = \int d^D k \, d^D l \, \frac{k^2 v}{N_l \mid_{\substack{q=0 \\ m_i=0}} (P_3 \mid_{m_3=0})^2 N_k}. \tag{9}$$

We shuffled UV-poles to artificial IR-poles. Usually, this would pose disturbing problems, but assuming some intermediate IR-regulator (Hadamard's Finite Part (HFP), small masses or DR itself) it is easy to see that in this special application the problematic terms drop out in the sum of the three graphs. Indeed, that this must be so is most easily seen by an analysis using HFP. The artificial IR-singularity is located at the submanifold $(l + k) = 0$, for example

$$HFP(\int d^D l \frac{1}{l^2(l+k)^4}) = \int d^D l(\frac{1}{l^2(l+k)^4} - \frac{\Theta(1 - \mid (l+k)^2 \mid)}{k^2(l+k)^4})$$
$$+ HFP(\int d^D l \frac{\Theta(1 - \mid (l+k)^2 \mid)}{k^2(l+k)^4}. \tag{10}$$

The first term in brackets has no artificial IR-singularity at $(l + k) = 0$, while the second term vanish in the sum of the graphs by using translation invariance of dimensional regularization and HFP. This means that we are allowed to shift $l \rightarrow l - k$, so that the $l$-integration decouples from $k$ and becomes

$$HFP(\int d^D l \frac{\Theta(1 - \mid l^2 \mid)}{k^2 l^4} = \frac{1}{k^2} \frac{2\pi^{D/2}}{\Gamma(D/2)} \frac{1}{(D - 4)}, \tag{11}$$

so that these terms factorize in separate $l$- and $k$-integrations. Accordingly, they vanish in the sum of all three graphs of table (1). A similar analysis can be carried out with any other IR-regulator. So the remaining contributions are UV-convergent in the first loop momentum, $l$ say. Now we isolate the UV-singularities of the second $k$-loop momentum:[1]

$$\frac{1}{N_k} = \frac{1}{N_k} \frac{(N_k \mid_{q=0} -N_k)^r}{N_k^r \mid_{q=0}} - \frac{1}{N_k^r \mid_{q=0}} \sum_{i=1}^{r} \binom{r}{i} (-N_k \mid_{q=0})^i (N_k \mid_{q=0})^{r-i}. \quad (12)$$

By an appropriate choice of $r$, determined by powercounting, all UV-divergences are isolated in the second term on the right-hand side. As UV-convergent terms do not contribute to our $Z$-factors, we now see that all remaining loop integrals can be solved by the fundamental identities Eq.(7,8). As the powers $\lambda_1, \lambda_2$ which are needed are completely determined by powercounting rules for the three graphs, we see that knowledge of the particle species, -which determines the powercounting-, and knowledge of the topology of the graph, -which determines the subgraphs-, is sufficient to calculate the $Z$-factor in MS-schemes. Using results on tensor integrals it is clear that this is true for any two-loop graph in a renormalizable theory.[1]

So let us collect the remaining contributing terms. There are terms which stem from the finite part of the subdivergences. These are the terms proportional to $d_0$ in Eq.(3). They multiply ordinary one-loop integrals. We stress that the finite coefficients $d_0$ stem from one-loop integrals evaluated at momentum transfer zero with vanishing internal masses. So they contain only one scale, the other loop momentum which flows through the subdivergence.

The only other surviving contribution evaluates the graph with all exterior momenta set to zero. Further, all masses belonging to the subdivergences are set to zero. This results again in one-scale integrals which can be obtained by applying Eq.(7,8) twice.

So our final approach to the calculation of $Z$-factors would be the following. First we determine the finite coefficients $d_0$, and also determine the true two-loop contributions.[1] We take the sum of the graph and all its subdivergences. We split the second loop integration in an UV finite and a divergent part. The UV finite part will not contribute in this sum. We keep in mind that we shift UV-divergences of the first loop integration to the final one. So we adjust our powercounting accordingly. This means that we choose the $r$ in Eq.(12) so that it renders even all the shifted expressions finite. Then, neither the first nor the second loop integration will produce UV-divergences for the terms generated by the first expression on the right-hand side of Eq.(12). So we can calculate these expessions in $D = 4$ dimensions, but then the difference between the terms generated by the counterterm graphs and the corresponding terms generated from the graph vanish. Here we assume that the graph is split into a finite integral representation and UV-divergent parts.[1]

So we have the remarkable result that only expressions generated by the second term on the right-hand side of Eq.(12) will contribute.

These terms have an unavoidable UV-divergence, so we do not insist on shifting UV-divergences of the subdivergence to the final loop integration. Rather we directly integrate out the first loop, which will result in powers of the second loop momentum. So we can do the remaining integration along the lines of Eq.(7,8), where one of the $\lambda$'s now has a non-integer value in $D \neq 4$ dimensions.

So the final approach avoids even partial integration methods and reduces to a book-keeping exercise in beta-functions. This gives us the $Z$-factors in MS-schemes. This result is only true for the sum of the graph plus its counterterm subgraphs.

We checked the method by applying it to massive two-loop QED and to $\phi^3$ in six dimensions and found total agreement with the standard results. Recent results on Z-factors in the full SM will be presented elsewhere.

## 3. Generalizations to other Renormalization Schemes

When we want to calculate $Z$-factors in schemes different from MS-schemes we can still apply this approach. Two changes have to be taken into account. On the one hand, the constants $d_0$ are clearly scheme dependent. And, on the other hand, we will need also the finite parts of the Laurent expansions in $(D-4)$. This involves the finite parts of the beta functions, but also the integral representations typically generated by massive two-loop integrals.[1,4,5,6] To obtain $Z$-factors, it is sufficient to know these finite parts at some fixed kinematical points (on-shell conditions, for example). Their evaluation at these points can be most easily obtained by using numerical approaches as described recently in the literature.[7]

## 4. Generalization to higher Loop Orders

We sketch briefly how this approach will generalize to the arbitrary loop level, but will present it in greater detail elsewhere. The crucial step is to identify all the maximal normal forests, according to Zimmermann's forest formula. These forests correspond to some lower order $Z$-factors, and we can reexpress these $Z$-factors by the corresponding forests, evaluated at zero momentum transfer and with vanishing internal masses. By shifting all UV-divergence to the last loop integration and isolating UV-divergences according to a generalization of Eq.(12) we can then expext similar cancellations as for the two-loop level, and only expessions accessible via Eq.(7,8) remain to be calculated, plus the terms generated by the $d_0$-type expressions, which are the finite parts of arbitrary massless two-point functions.

So what has to be shown is that one can obtain these arbitrary massless two-point functions in general; that the forest formula gives just the right structure to allow for all the cross cancellations which cancel non-local terms; and that a tool exists which locates UV-divergences in the last loop-integration. Partial integration methods fail at the higher loop level, but a combination of HFP and the forest formula is successful at this point, as will be shown elsewhere.

## 5. $\phi^3$ in six Dimensions

To give a more concrete example, let us have a look at $\phi^3$ theory in six dimensions. So we have $D = 6 - 2\varepsilon$, and our normalization is such that we have

$$-\bigcirc- = \int d^D l \, d^D k \, \frac{1}{(l^2 - m^2)((l+q)^2 - m^2)((l+k)^2 - m^2)}$$

$$\frac{1}{(k^2 - m^2)((k-q)^2 - m^2)}. \tag{13}$$

According to the previous section, the calculation of the mass renormalization constant in the MS-scheme $Z_m$ reduces to expressions

 $\displaystyle = \int d^D l\, d^D k\, \frac{1}{(l^2)^2(l+k)^2(k^2-m^2)^2}$

$\displaystyle \qquad + \int d^D l\, d^D k\, \frac{1}{(k^2)^2(l+k)^2(l^2-m^2)^2} + \text{other terms},$

 $\displaystyle = \int d^D l\, d^D k\, \frac{(k^2)^\varepsilon}{(l^2)^2(l+k)^2(k^2-m^2)^2} + \text{other terms},$

 $\displaystyle = \int d^D l\, d^D k\, \frac{(l^2)^\varepsilon}{(l^2)^2(l+k)^2(k^2-m^2)^2} + \text{other terms}, \tag{14}$

where "other terms" takes into account either finite terms or terms which vanish when one sums over the three graphs.

We can calculate the wave function renormalization $Z_\phi$ along similar lines (first differentiate with respect to $q^2$ and then set $q^2 = 0$ in the second loop integration, or calculate with a vanishing mass $m^2 = 0$, and use $q^2$ as the remaining scale in the last integration).

All the above integrals can be solved by using Eq.(7,8), and one easily finds the standard result:[8]

$$\begin{aligned}
Z_\phi &= \frac{1}{6}\frac{1}{(D-6)^2} + \frac{1}{3}\frac{1}{D-6}, \\
Z_m &= m^2\left(-\frac{1}{(D-6)^2} - \frac{1}{2}\frac{1}{D-6}\right).
\end{aligned} \tag{15}$$

## 6. Conclusions

We briefly sketched an approach to $Z$-factors in the MS-scheme, which reduces the calculation of these $Z$-factors to a book-keeping exercise in beta-functions. The method applies to arbitrary processes in the SM, and is of particular value there; it most easily overcomes the problems with complicated analytic expressions in intermediate steps, typically generated by loop integrals involving various different masses.

### Acknowledgements

It is a pleasure to thank R. Delbourgo and P. Jarvis for many stimulating discussions and interest in this work. I would like to thank the organizers of this conference for their kind hospitality and support.

This work was supported under grant A69231484 from the Australian Research Council.